\documentclass[a4paper,english,11pt]{article}
\usepackage[dvips]{graphicx,epsfig,color}
\usepackage{wrapfig,rotating}
\usepackage{amssymb,amsmath,array}
\usepackage{subfigure}
\usepackage{wrapfig}
\usepackage{sidecap}

\begin{document}
\title{Description of soft diffraction in the framework of reggeon 
calculus. Predictions for LHC}

\author{{\slshape Alexei Kaidalov$^{1}$, Martin Poghosyan$^{2}$}\\[1ex]
$^{1}$ Institute of Theoretical and Experimental Physics, 117526 Moscow, Russia\\
$^{2}$ Universit\`a di Torino and INFN, 10125 Torino, Italy
}

\maketitle

\begin{abstract}
A model, based on Gribov's Reggeon calculus, is proposed and applied to 
processes of soft diffraction at high energies. It is shown that by 
accounting for absorptive corrections for all legs of triple-Regge and 
loop diagrams a good description of experimental data on inelastic soft 
diffraction can be obtained. In this paper we give a brief description 
of the model and of its predictions for LHC energies.
\end{abstract}
\section{Introduction}
\vspace{-7pt}
The process of soft single- and double- diffraction dissociation are closely related to small 
angle elastic scattering in which each of the incoming hadrons may become a system which will 
then decay into a number of stable final state particles.
Regge-pole theory is the main method for description of high-energy soft processes. 
In this approach (see \cite{Kaidalov_PRP}), the inclusive cross-section of single and double diffraction 
dissociations is described by triple-Reggeon and loop diagrams, respectively.
Triple-Reggeon description is in good agreement with the FNAL and ISR data for 
soft diffraction dissociation \cite{LowEnFit}. However, the higher-energy data 
from SPS and Tevatron do not show the increase of the cross section with energy 
expected from the simple fits and the contribution of triple-Pomeron vertex (in 
the elastic scattering amplitude) violates unitarity. A number of different 
approaches have been proposed in order to be in agreement with the data from 
higher energy experiments: non-gaussian parameterization for Reggeon-hadron vertex 
\cite{NonGaussRes}, renormalization \cite{PomFluxRen} or damping \cite{PomFluxDump} 
of the Pomeron flux. A more realistic approach suggested in \cite{GostmanEtAl} and 
\cite{LunaEtAl} by the inclusion of initial state elastic scattering corrections to 
the  triple-Reggeon vertices. However, the analysis done in \cite{Martynov} shows 
that this correction is not enough for restoring the $s$-channel unitarity.\\
Besides of its own role, the theoretical knowledge of soft diffraction is also 
important in analyzing hard diffraction data, which is an active area of study 
at HERA and Tevatron and will continue to be interesting at LHC. The knowledge 
of the interaction between Pomerons (enhanced diagrams) is important for analyzing 
data at very high energies, too. The contribution of these diagrams can be essential in 
hadron-nucleus and especially in nucleus-nucleus collisions, where the thermalisiation 
and the quark-gluon plasma formation strongly depend on the strength of interactions 
between Pomerons \cite{KaidalovNuclPhysA525_39c}.\\
%
In this article we propose to describe data on soft diffraction dissociation in $pp$ and 
$p\bar{p}$ interactions taking into account all possible non-enhanced absorptive corrections 
to triple-Regge vertices and loop diagrams. This approach describes available 
data on high-mass soft diffraction in the energy range from ISR, FNAL to Tevatron.
The article is organized as following: 
In the next two sections we briefly describe the Regge-pole approach, Gribovs' Reggeon calculus and 
AGK cutting rules. Our proposed model is presented in Section 4 and its predictions are compared with data
in Section 5. 
%
\vspace{-15pt}
\section{Single Regge-pole approximation}
\vspace{-10pt}
In Regge theory, the simplest singularity in the $j$-plane is a moving pole $\alpha(t)$  in the
$t$-channel (see the leftmost grah in Fig.~\ref{Fig:MultiReggeExchange}) and in the small $t$-region 
the scattering amplitude, $M(s,t)$, of the process $a+b \rightarrow c+d $ can be parameterized as:
\vspace{-9pt}
\begin{equation}
M(s,t)=\gamma (0) \eta( \alpha(0) )(s/s_{0})^{\alpha (0)-1} \exp(\lambda(s)t).
\label{eq:OneReggeMpar}
\vspace{-5pt}
\end{equation}
Here $\eta(\alpha(t))$ is the signature factor, $\gamma(t) \equiv g_{ac}(t)g_{bd}(t)$ is the
factorization residue, $\lambda(s)=R^{2}+\alpha_{R}^{\prime}~ln(s/s_{0})$.
The  parameter $R^{2}$ characterizes the $t$-dependence of the product of residue function and 
of the signature factor. In our notations the normalization of the scattering amplitude is such that
$\sigma_{tot}=8\pi ImM(s,0)$ and $d\sigma_{el}/dt = 4\pi |M(s,t)|^{2}$.\\
Because the Pomeron's intercept is larger than unity (which is required in order to 
guarantee the growth of the total cross-section: $\sigma_{tot}\sim s^{\Delta}$, 
$\Delta \equiv \alpha_{P}(0)-1$), 
the corresponding cross-section grows as a power function of $s$
and therefore the contribution of the Pomeron-pole in the scattering amplitude violates unitarity. 
The easiest way to 
restore the unitarity is to take into account branch points which correspond to 
multi-Reggeon exchange. The calculation of the multi-Reggeon exchange amplitude is 
possible in eikonal (or eikonal-like) approximation, where only single particle 
intermediate states are taken into account.
\vspace{-15pt}
\section{Eikonal approximation and AGK cutting rules}
\begin{SCfigure}[][h]
  \centering
\includegraphics[width=0.5\textwidth]{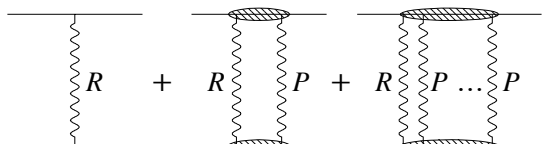}
  \caption{
Single pole and $RP^{n}$ cut contribution in the elastic scattering amplitude.
$R$ stands for secondary Reggeon and for Pomeron.
}
\label{Fig:MultiReggeExchange}
\vspace{-12pt}
\end{SCfigure}

\vspace{-10pt}
Regge poles are not the only singularities in the complex
angular momentum plane. Exchange of several Reggeons in
$t$-channel leads to moving branch points in the $j$-plane (Fig.~\ref{Fig:MultiReggeExchange}).
A Regge pole exchange can be interpreted as corresponding
to single scattering while Regge cuts correspond to multiple
scatterings on constituents of hadrons.
In case of ”supercritical” Pomeron~($\Delta >0$) the contribution of $n$-Pomeron exchange in 
the scattering amplitude ($M^{(n)}_P(s, 0)\sim s^{n\Delta}$) is increasing with the increase of 
the energy and the entire series of $n$-Pomeron exchange should be summed.
\vspace{-6pt}
On the contrary, 
the contribution of the branch points concerned with the exchange of several secondary-Reggeons 
decreases very quickly with increasing collision energy and the contribution of such branch points 
can be neglected with respect to the branch points due to the exchange of one secondary Reggeon 
and Pomerons that are needed for properly matching low energy data.
For instance, in eikonal approximation the amplitude of $n$-Pomeron exchange can be written in the following 
form \cite{TMreggeclac}:
\vspace{-6pt} 
\begin{equation}
M^{(n)}(s,t)=\frac{(2i)^{n-1}}{n!}
\int
\prod_{i=1}^{n} 
\left[
M^{(1)} (s,{\mathbf q}_{i\bot}^{2})
\frac{d^{2} {\mathbf q}_{i\bot}}{2\pi} 
\right]
\delta \left(  {\mathbf q}_{\bot} - \sum_{i=1}^{n} {\mathbf q}_{i \bot} \right).
\label{eq:NReggeM}
\vspace{-7pt}
\end{equation}
Using the parameterization (\ref{eq:OneReggeMpar}) for the Regge pole contribution and performing
the integration over the transverse momenta of Reggeons in Eq. (\ref{eq:NReggeM})
it can be shown that the account of the multi-Pomeron exchanges results in the unitarization of 
the scattering amplitude,  which leads to the Froissart behavior of the total cross section for
 $s \gg m_{N}^{2}$: $\sigma_{tot} \simeq 8\pi \alpha_{P}^{\prime} \Delta \ln ^{2} (s/s_{0})$.\\
%
In the language of Regge poles the multiparticle production processes are related to 
cut-Reggeon diagrams. Abramovski, Kancheli and Gribov (AGK) proposed rules \cite{AGK}
for calculating the discontinuity of the matrix element that represent the generalization 
of the optical theorem for the case of multi-Pomeron exchange.
The basic results of AGK needed for the following discussion are:
a) There is one and only one cut-plane which separates the initial and final states of the scattering.
b) Each cut-pomeron gives an extra factor of ($-2$) due to the discontinuity of the pomeron amplitude. 
c) Each un-cut pomeron obtains an extra factor of 2 since it can be placed on both sides of the cut-plane. 
\vspace{-15pt}
\section{The Model}
\begin{SCfigure}[][h]
  \centering
\includegraphics[width=0.32\textwidth]{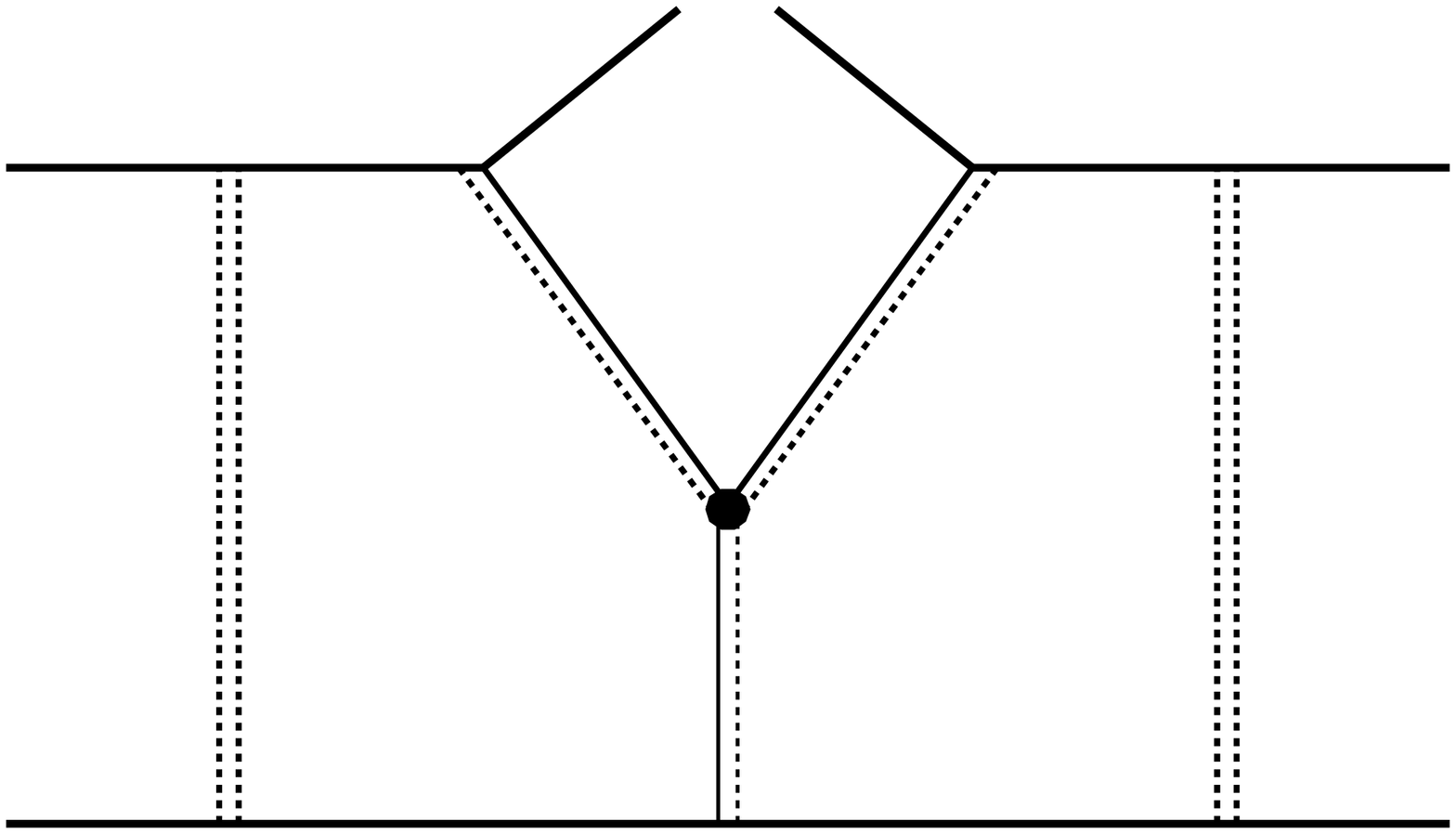}
\includegraphics[width=0.32\textwidth]{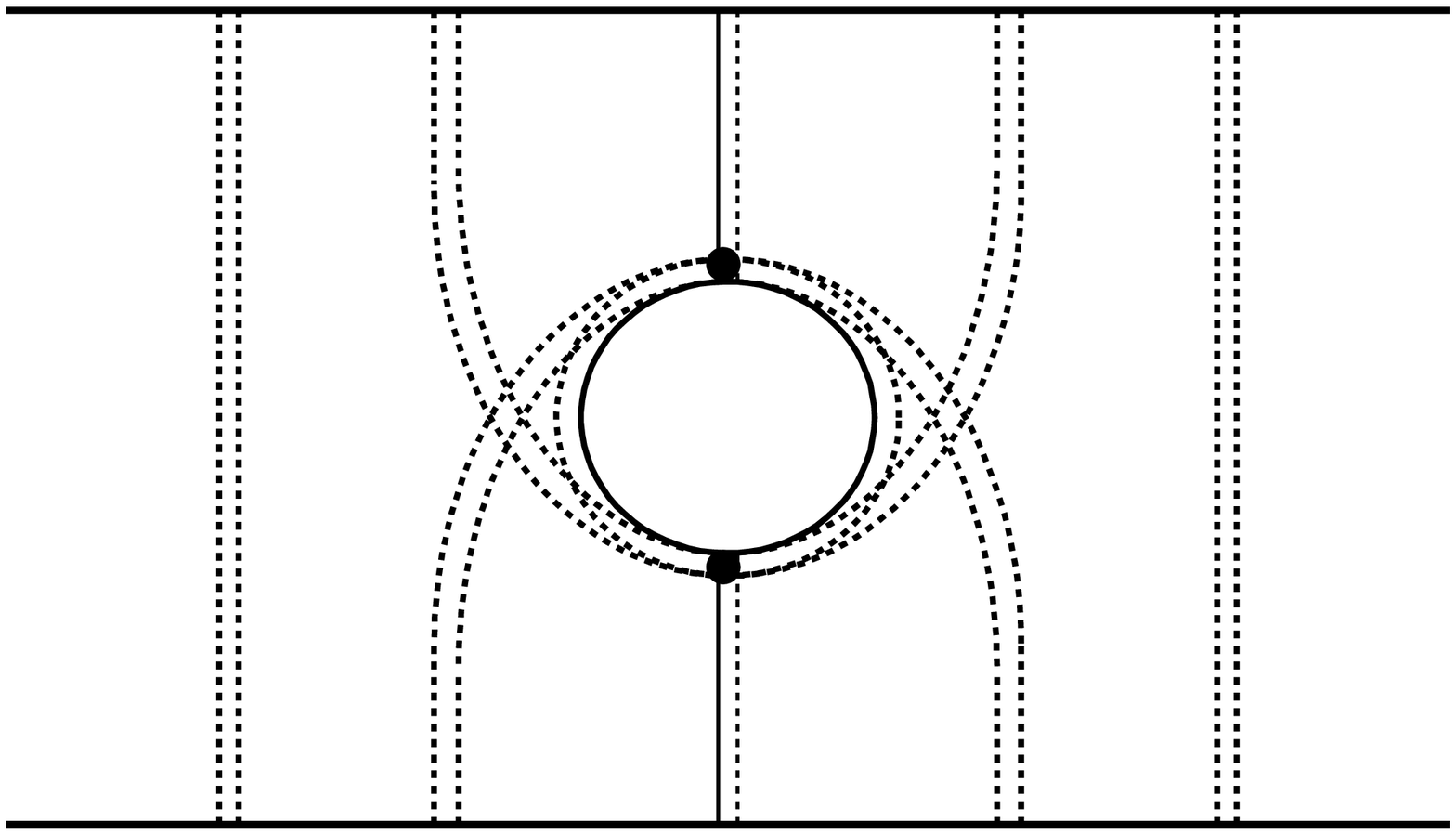}
  \caption{ Eikonalised triple-Reggeon and loop diagrams which are proposed to
   describe single- and double- diffractive processes in hadron-hadron collisions.}
\label{Fig:DrGraph}
\vspace{-10pt}
\end{SCfigure}

\vspace{-5pt}
We propose to describe single- and double- diffraction processes by such diagrams where any 
number of Pomeron exchanges is taken into account together with each $R$ of the triple-Reggeon 
and loop diagrams and as well as the screening corrections are 
considered, as shown in Fig.~\ref{Fig:DrGraph}.   
In this Figure the solid 
line accompanied with a dashed line corresponds to one Reggeon 
(Pomeron or secondary-Reggeon) exchange together with any 
number Pomeron exchange. The double-dashed lines stand for 
eikonal screening.\\
The theory does not give any prediction on the structure of the vertices for $n$ 
Pomeron to $m$ Pomeron transitions. The simplest approximation is to assume an 
eikonal-type structure. In this approximation the general approach of constructing 
elastic scattering amplitude with account of enhanced diagrams has been proposed in 
\cite{KT-MP}.  Assuming $\pi$-meson exchange dominance of multi-Pomeron interaction 
vertices, the authors summed high order enhanced diagrams iterating multi-Pomeron 
vertices in both, $s$- and $t$- channels. In that article it was demonstrated that 
the inclusion of these diagrams in most of the cases leads to predictions that are 
very close to the results of eikonal type models, where a Pomeron with suitably 
renormalized intercept is used.\\
For calculating the diagrams shown in Fig.~\ref{Fig:DrGraph} we assume the mentioned 
$\pi$-meson exchange dominance of multi-Pomeron interaction vertex corresponding to 
the transition of $n$ Pomerons to $m$ Pomerons:
\vspace{-7pt}
\begin{equation}
\lambda^{(n,m)}=r_{3P} g_{\pi}^{n+m-3} \exp \left( -R_{\pi}^{2} \sum_{i=1}^{n+m}q_{i}^{2} \right).
\label{eq:PomPomVrt}
\vspace{-3pt}
\end{equation}
As a secondary Regge pole we consider $f$-trajectory. The conservation laws allow us to 
assume the same pion dominance at the same transition with participation of $f$-trajectory.
In these terms according to the AGK cutting rules, the cross-section 
corresponding to the cut dressed triple-Reggeon graph for the process $a+b \rightarrow X +b$ has 
the form\footnote{If the transferred momentum is very low and the mass of the 
diffracted system is high the $\pi$-meson exchange plays an important role. This we take into 
account based on the OPER model \cite{OPER}.}:
\vspace{-3pt}
\begin{equation}
\frac{d \sigma}{d\zeta }=
\frac{1}{2}
\sum_{i,j,k = P,R} G_{ijk}
\int d{\bf b} d{\bf b_{1}}
\Gamma^{i}_{b \pi}(\zeta_{2}, {\bf b_{2}})
\Gamma^{j}_{b \pi}(\zeta_{2}, {\bf b_{2}})
\Gamma^{k}_{a \pi}(\zeta, {\bf b_{1}})
\exp\{-2\Omega_{ab}(\xi, {\bf b})\}
\label{eq:sigmaDr3R}
\vspace{-3pt}
\end{equation}
Here we introduce the following notations
\vspace{-6pt}
\begin{eqnarray}
\zeta=ln(M^{2}_{X}/s_{0}), \,  \, \zeta_{2}=\xi -\zeta,  \,  \, {\bf b_{2} = b - b_{1}}, \,\,
\Omega_{\alpha \beta }(\zeta, {\bf b}) = \frac{g_{\alpha \beta}}{\lambda_{\alpha \beta}}
\exp \left\lbrace \Delta \zeta - \frac{{\bf b}^{2}}{4 \lambda_{\alpha \beta}} \right\rbrace,
\label{eq:sigma3R1}\\
\Gamma^{P}_{\alpha \beta }(\zeta, {\bf b}) = 1-e^{-\Omega_{\alpha \beta }(\zeta, {\bf b})},  \,  \,
\Gamma^{R}_{\alpha \beta }(\zeta, {\bf b}) = \frac{g_{\alpha \beta }^{R}}{\lambda_{\alpha \beta}^{R}}
\exp \left\lbrace (\alpha_{R}-1)\zeta - \frac{{\bf b}^{2}}{4\lambda_{\alpha \beta}^{R}} -\Omega_{\alpha \beta }(\zeta, {\bf b}) 
\right\rbrace.  
\nonumber
\vspace{-15pt}
\end{eqnarray}
$G_{ijk}$ stand for triple-Regge vertices strength. The expression 
of the cross-section in the ($\zeta , t$)-space is rather long and we do 
not present it here.\\
Analogously can be calculated the cross-section corresponding to the cut dressed loop diagram standing for the process 
$a+b \rightarrow X_{1} + X_{2}$ and it has the following form:
\vspace{-5pt}
\begin{eqnarray}
\frac{d\sigma}{d \zeta_{1}d\zeta_{2}}=
\frac{1}{4}
\sum_{i,j,k,l = P,R}G_{ijk}G_{lik}
\int d \mathbf{b} d\mathbf{b_{1}}d\mathbf{b_{2}}
\Gamma_{\pi a}^{i}(\zeta_{1},\mathbf{b_{1}})
\Gamma_{\pi b}^{l}(\zeta_{2},\mathbf{b_{2}})
\Gamma_{\pi \pi}^{j}(\zeta_{3},\mathbf{b_{3}})
\Gamma_{\pi \pi}^{k}(\zeta_{3},\mathbf{b_{3}}) \nonumber\\
\times \exp \{-2\Omega_{ab}(\xi, {\bf b})
-2\Omega_{a\pi}(\xi - \zeta_{1}, {\bf b- b_{1}})
-2\Omega_{b\pi}(\xi - \zeta_{2}, {\bf b- b_{2}})\}
\label{eq:sigmaDD}
\end{eqnarray}
Here in addition to (\ref{eq:sigma3R1}) we used the following notations:
\vspace{-6pt}
\begin{eqnarray}
\zeta_{1}=ln(M^{2}_{X_{1}}/s_{0}), \,  \,
\zeta_{2}=ln(M^{2}_{X_{2}}/s_{0}), \,  \,
\zeta_{3}=\xi - \zeta_{1}-\zeta_{2}, \, \, {\bf b_{3} = b - b_{1} -b_{2}}\nonumber
\end{eqnarray}
\newline
\vspace{-45pt}
\section{Extraction of the parameters from experimental data}
\begin{figure*}[h!]
\vspace{-15pt}
\centering{
\begin{minipage}[l]{0.47\textwidth}
\includegraphics[width=1.\textwidth]{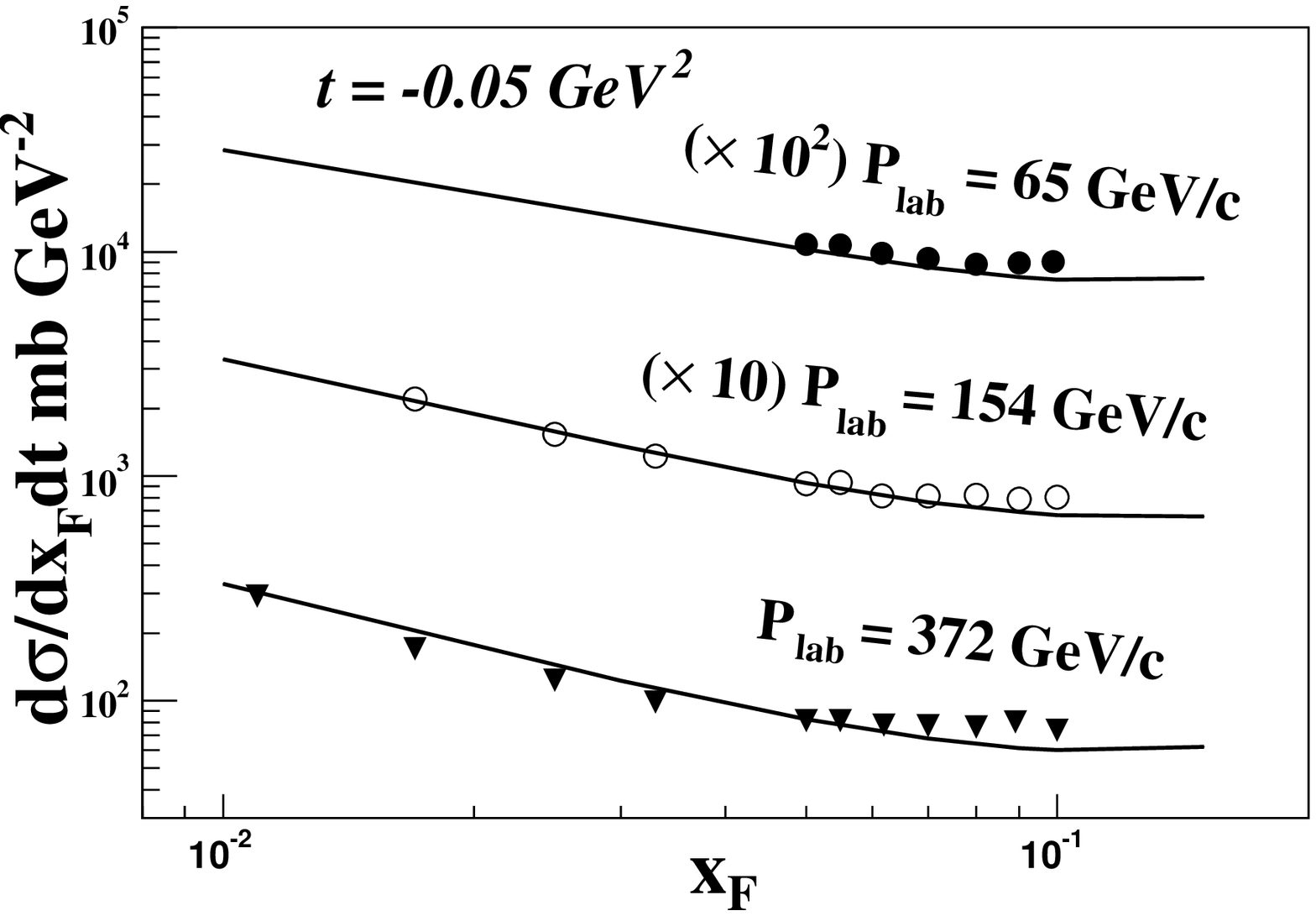}
\end{minipage}
\begin{minipage}[l]{0.47\textwidth}
\includegraphics[width=1.\textwidth]{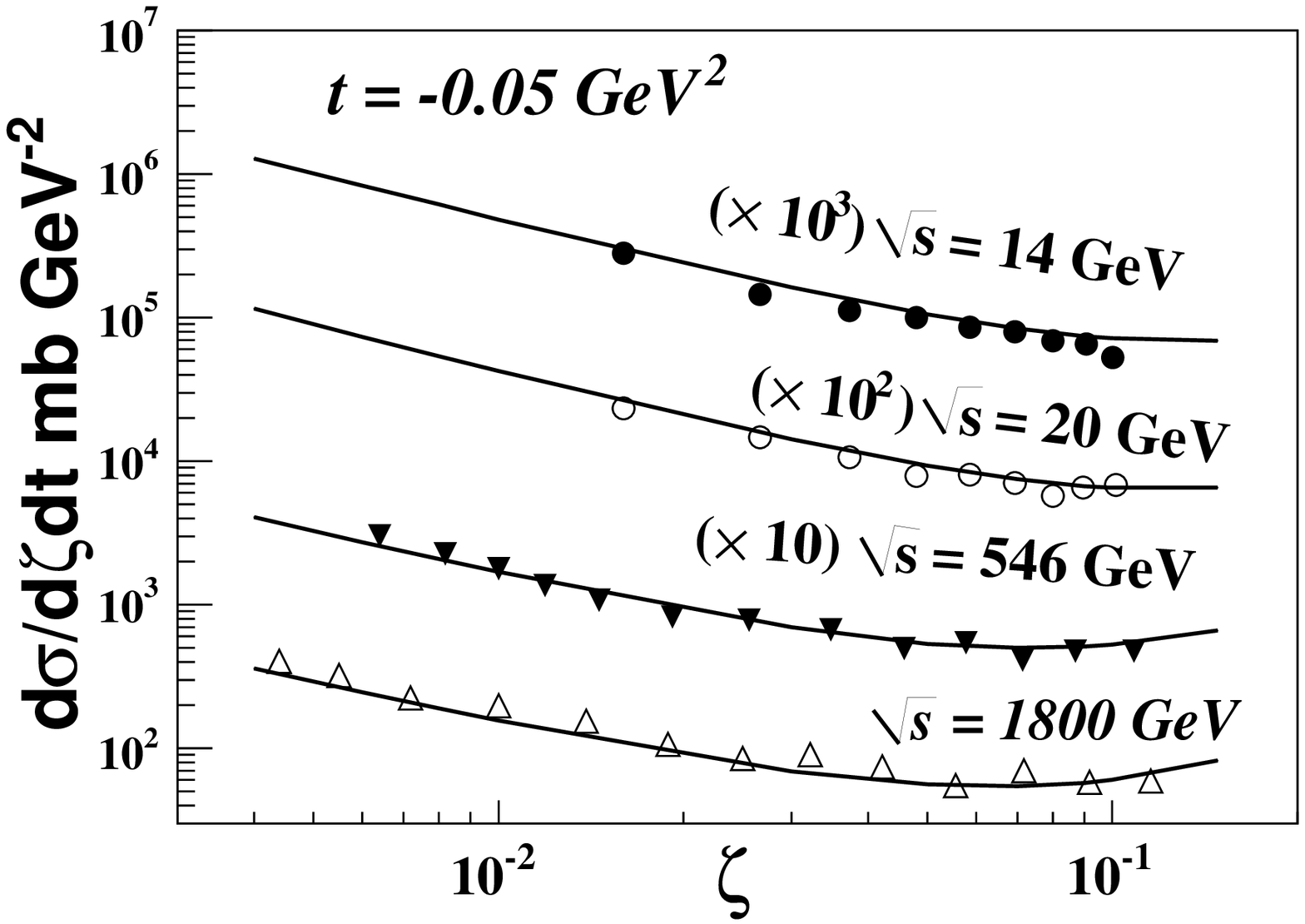}
\end{minipage}
}
\caption{
Double differential cross-section $d^{2}\sigma/d\zeta dt$ for $p(\bar{p})+p \rightarrow p(\bar{p})+X$ 
measured at Fermilab at various $\sqrt{s}$ and fixed $t$. The 
data are taken from \cite{PomFluxRen, Akimov}.}
\label{Fig:Comp1}
\vspace{-10pt}
\end{figure*}

\begin{figure*}[h!]
\vspace{-10pt}
\centering{
\begin{minipage}[l]{0.455\textwidth}
\includegraphics[width=1.\textwidth]{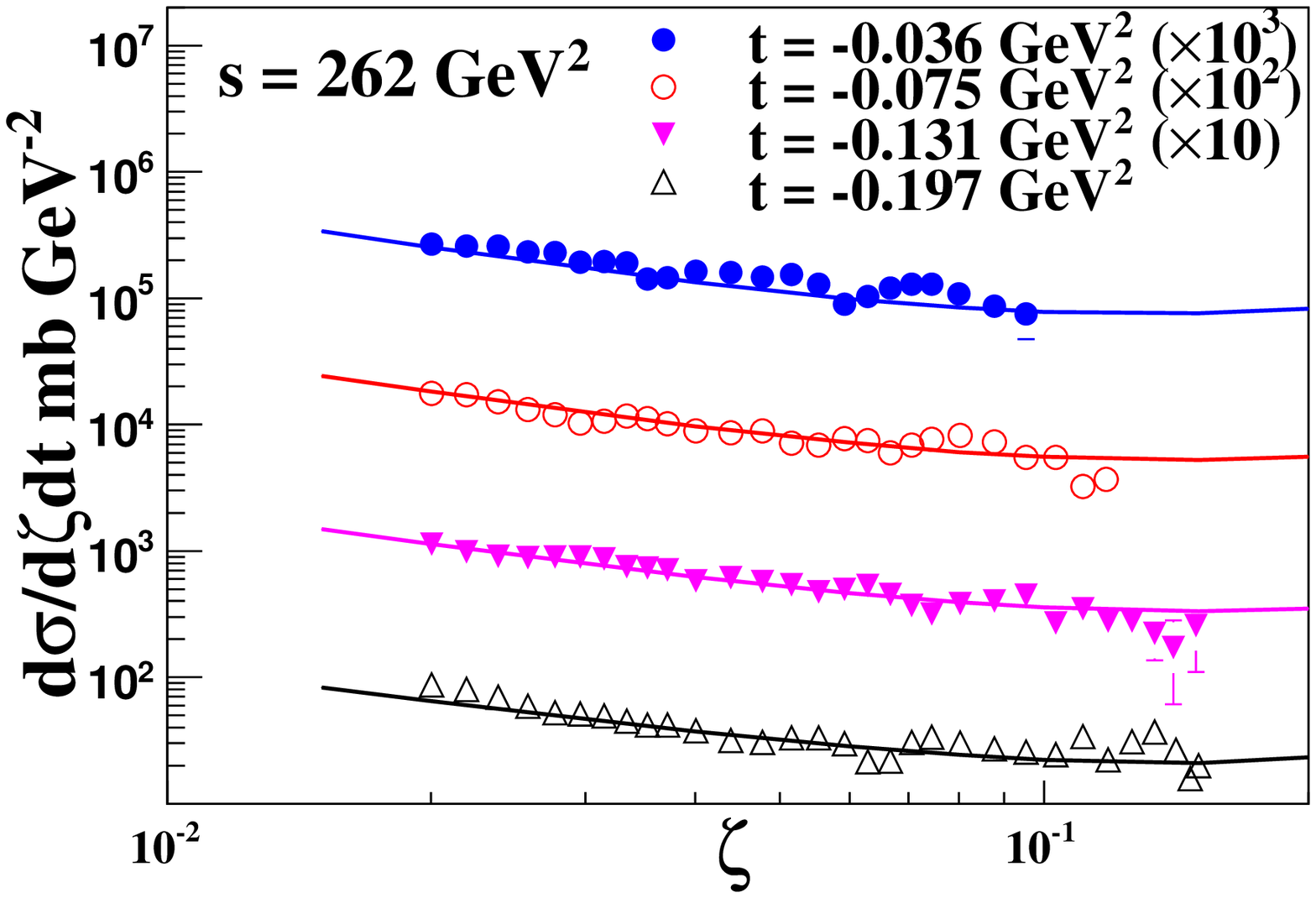}
\end{minipage}
\begin{minipage}[l]{0.455\textwidth}
\includegraphics[width=1.\textwidth]{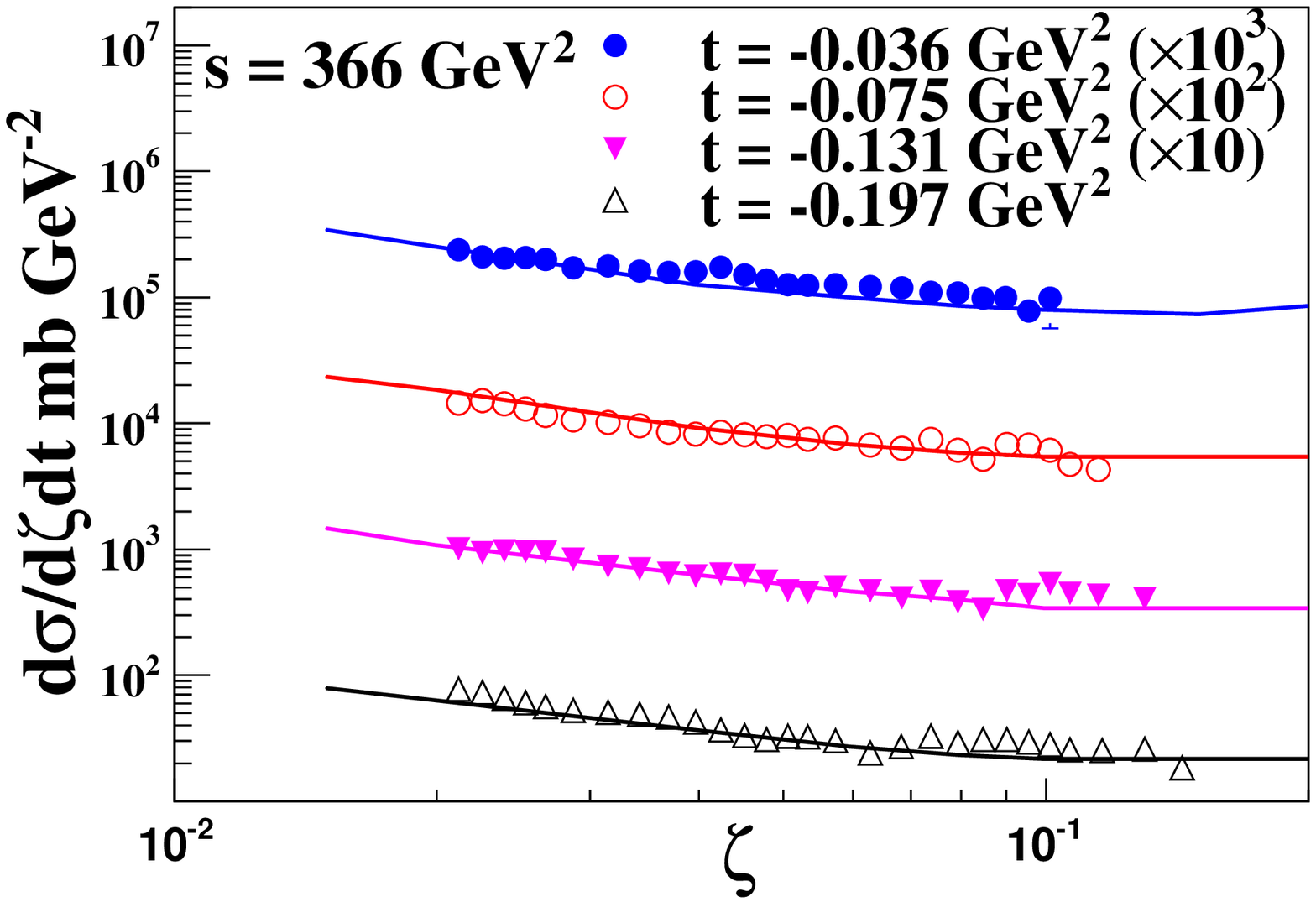}
\end{minipage}
\vspace{-5pt}
\begin{minipage}[l]{0.455\textwidth}
\includegraphics[width=1.\textwidth]{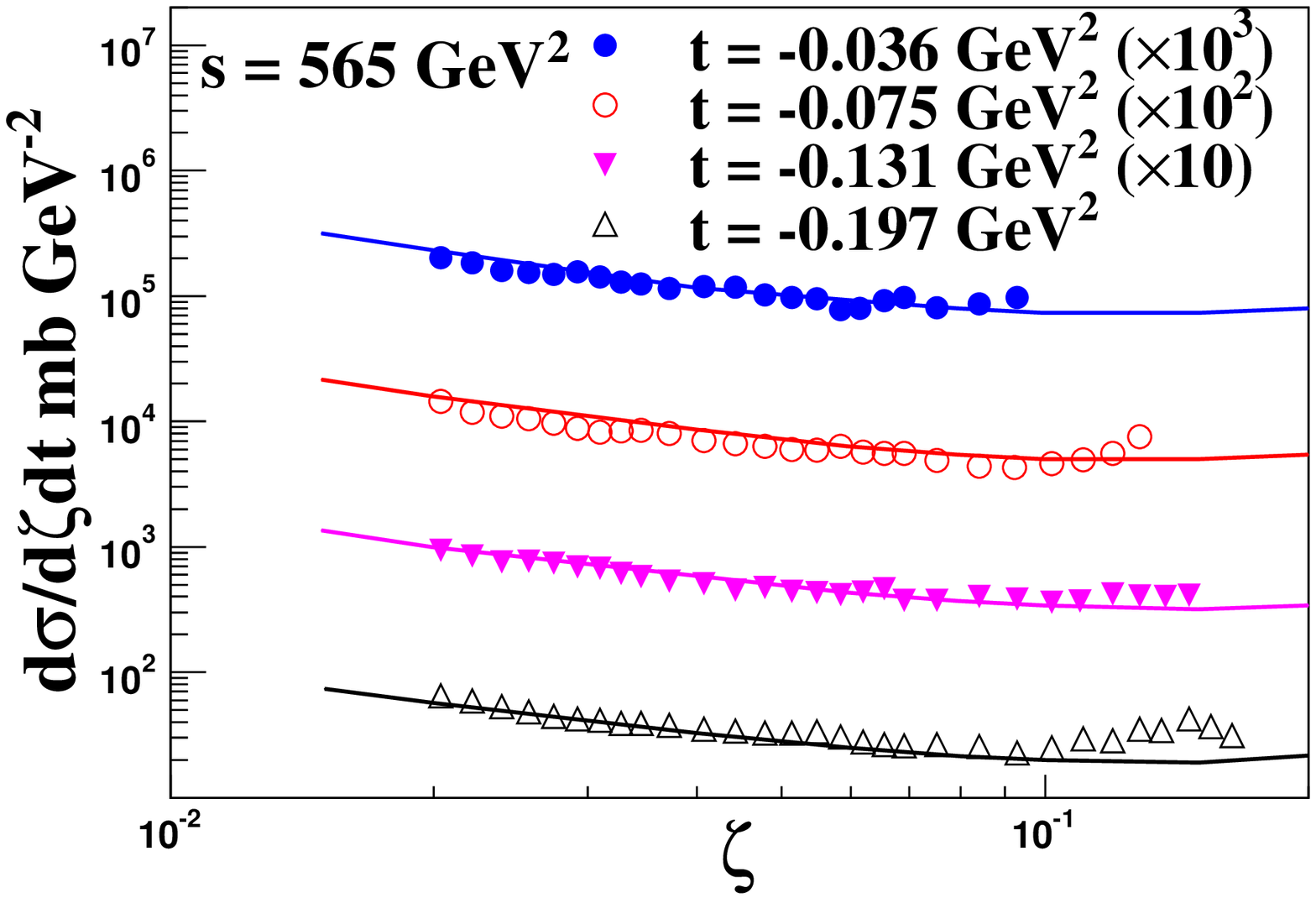}
\end{minipage}
\begin{minipage}[l]{0.455\textwidth}
\includegraphics[width=1.\textwidth]{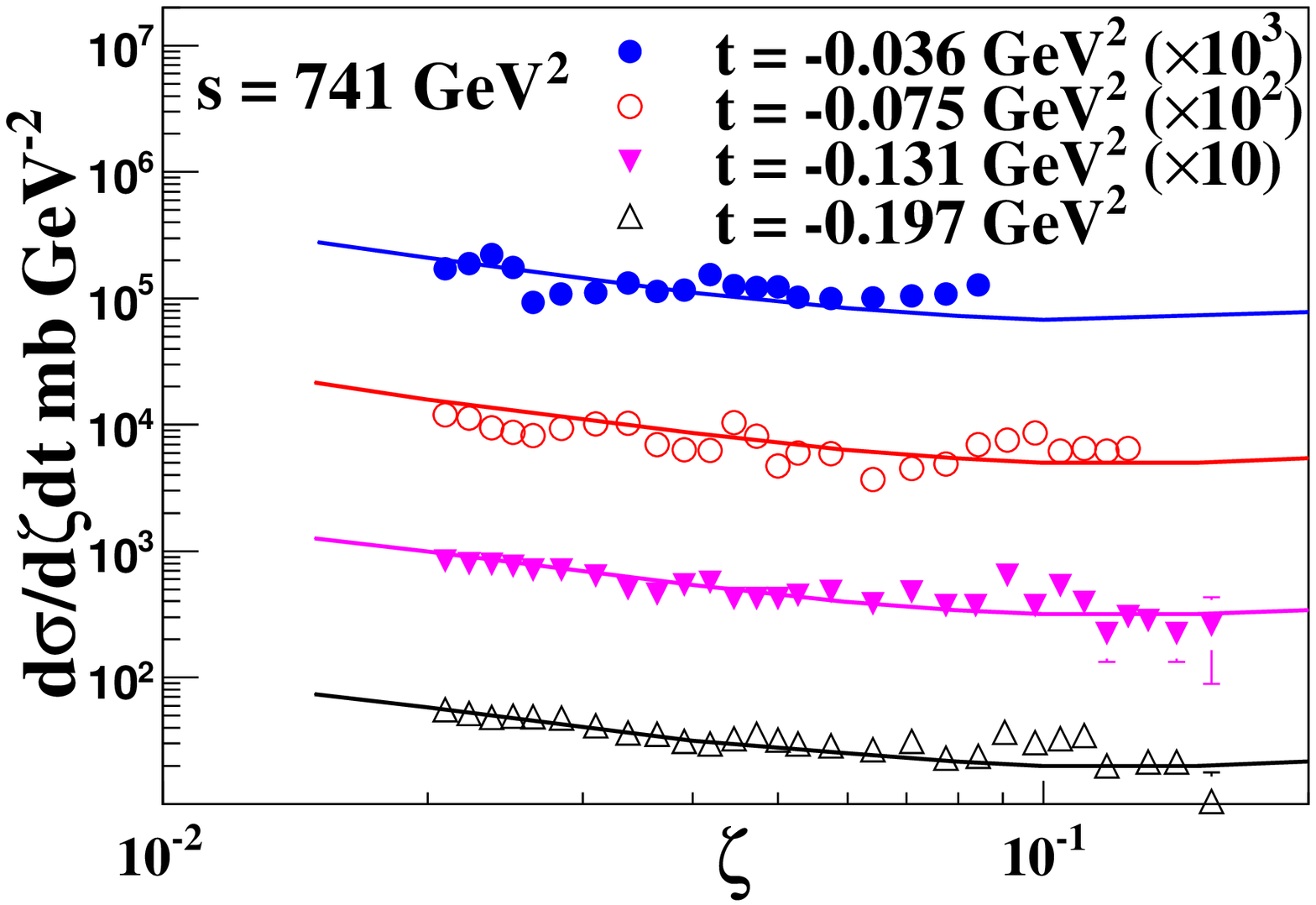}
\end{minipage}
}
\caption{ 
Double differential cross-section $d^{2}\sigma/d\zeta dt$ for $pp \rightarrow pX$ 
measured at Fermilab at various $\sqrt{s}$ and t.  The data are taken from \cite{Schamberger}.}
\label{Fig:Comp2}
\end{figure*}

\begin{figure*}[h!]
\vspace{-10pt}
\centering{
\begin{minipage}[l]{0.455\textwidth}
\includegraphics[width=1.\textwidth]{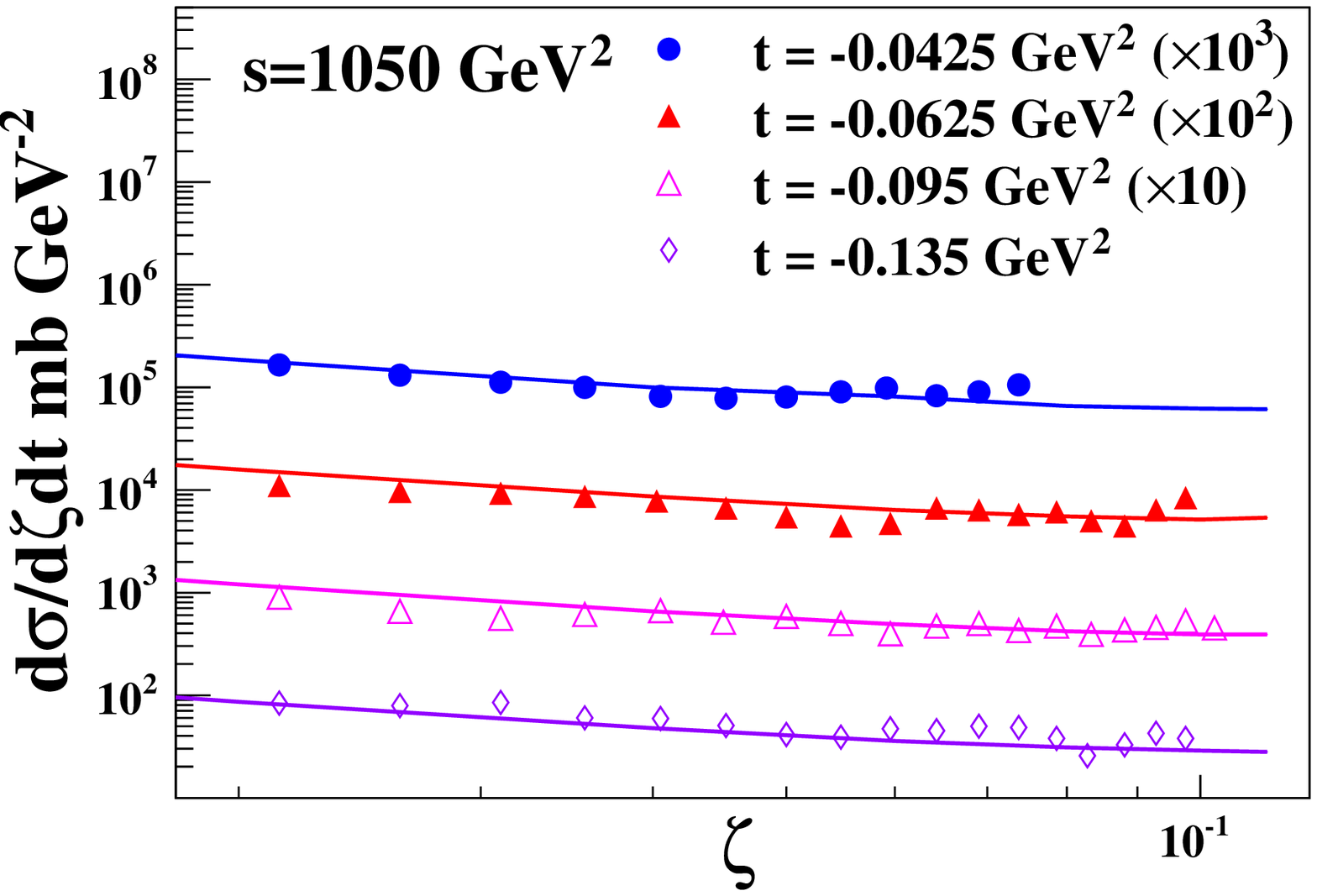}
\end{minipage}
\begin{minipage}[l]{0.455\textwidth}
\includegraphics[width=1.\textwidth]{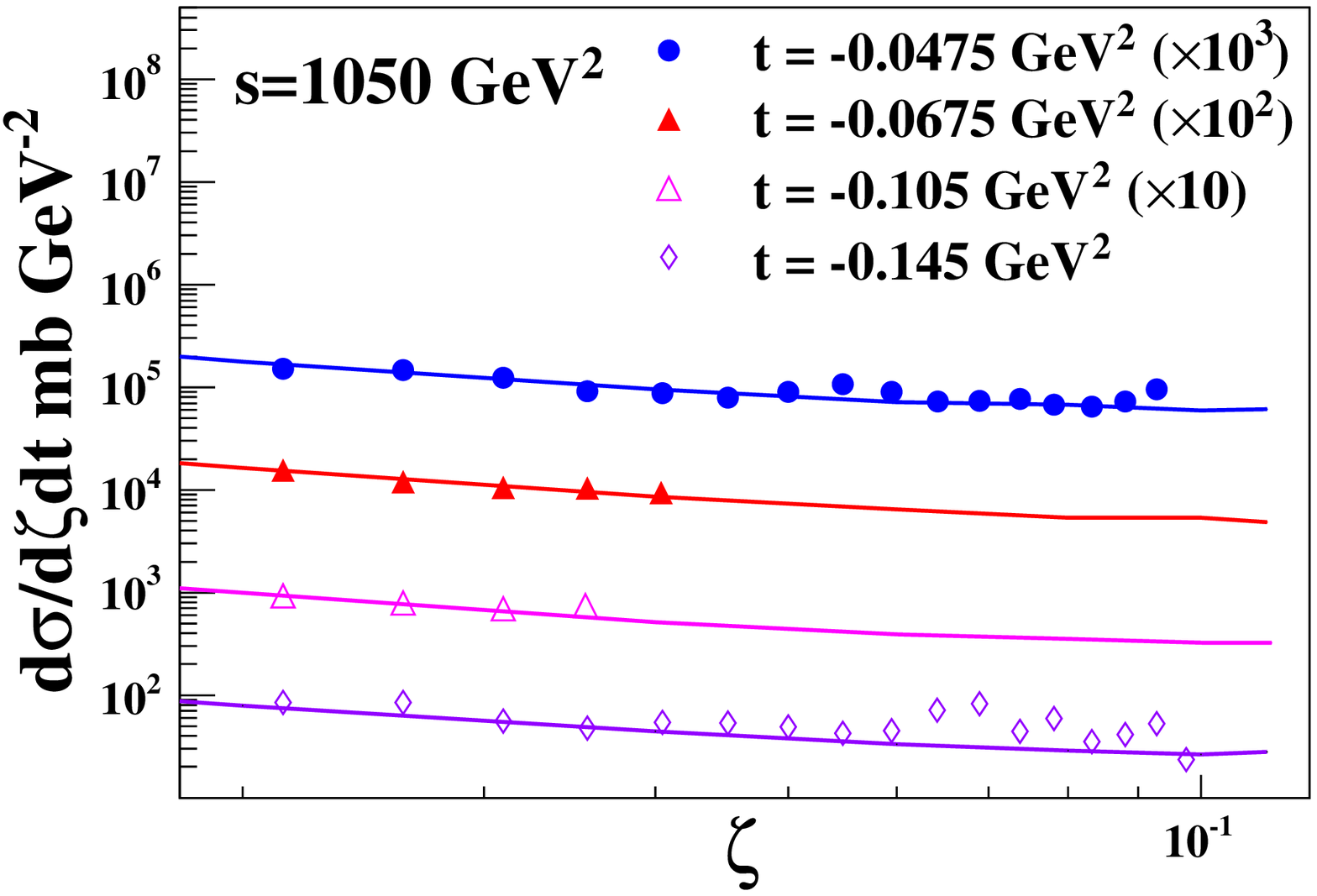}
\end{minipage}
\begin{minipage}[l]{0.455\textwidth}
\includegraphics[width=1.\textwidth]{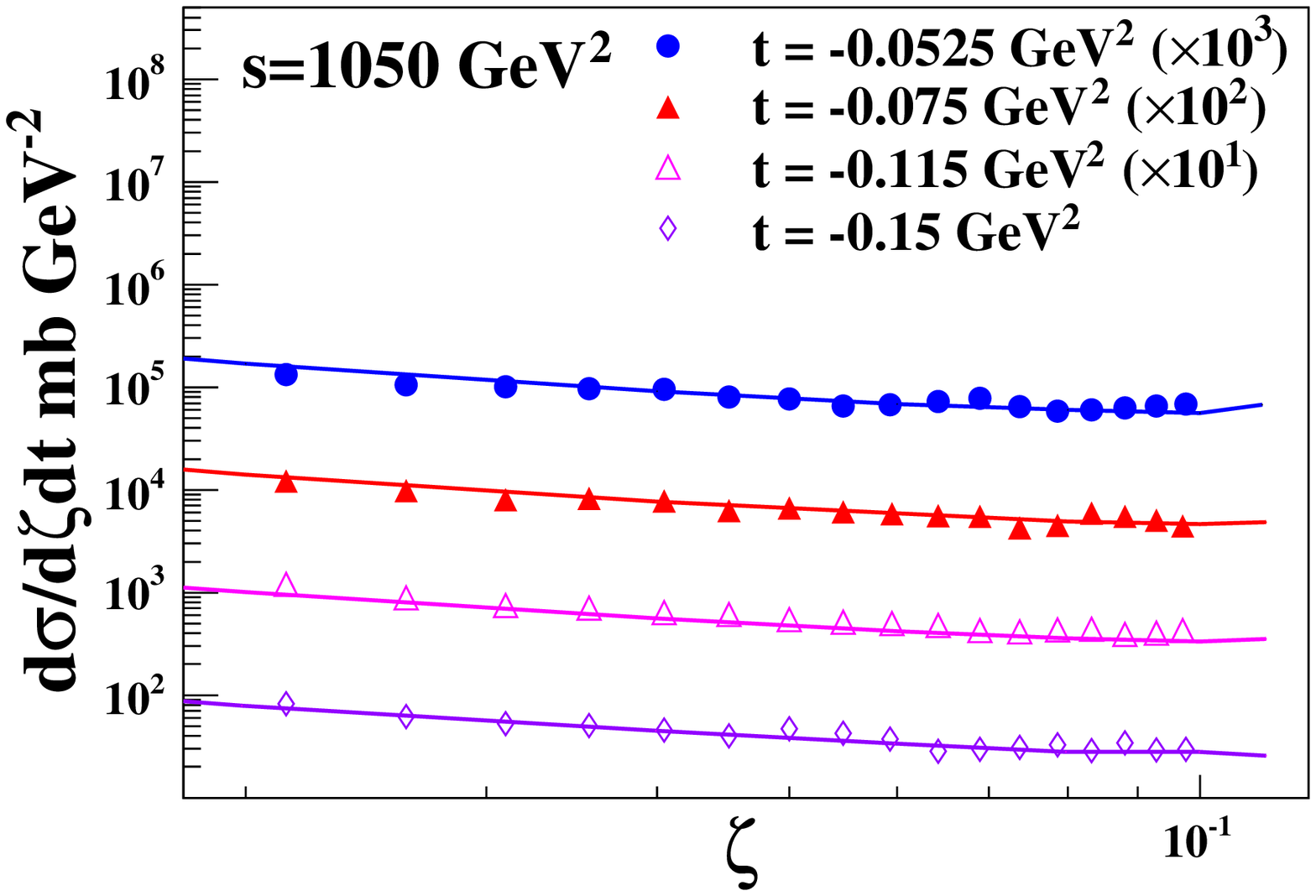}
\end{minipage}
\begin{minipage}[l]{0.455\textwidth}
\includegraphics[width=1.\textwidth]{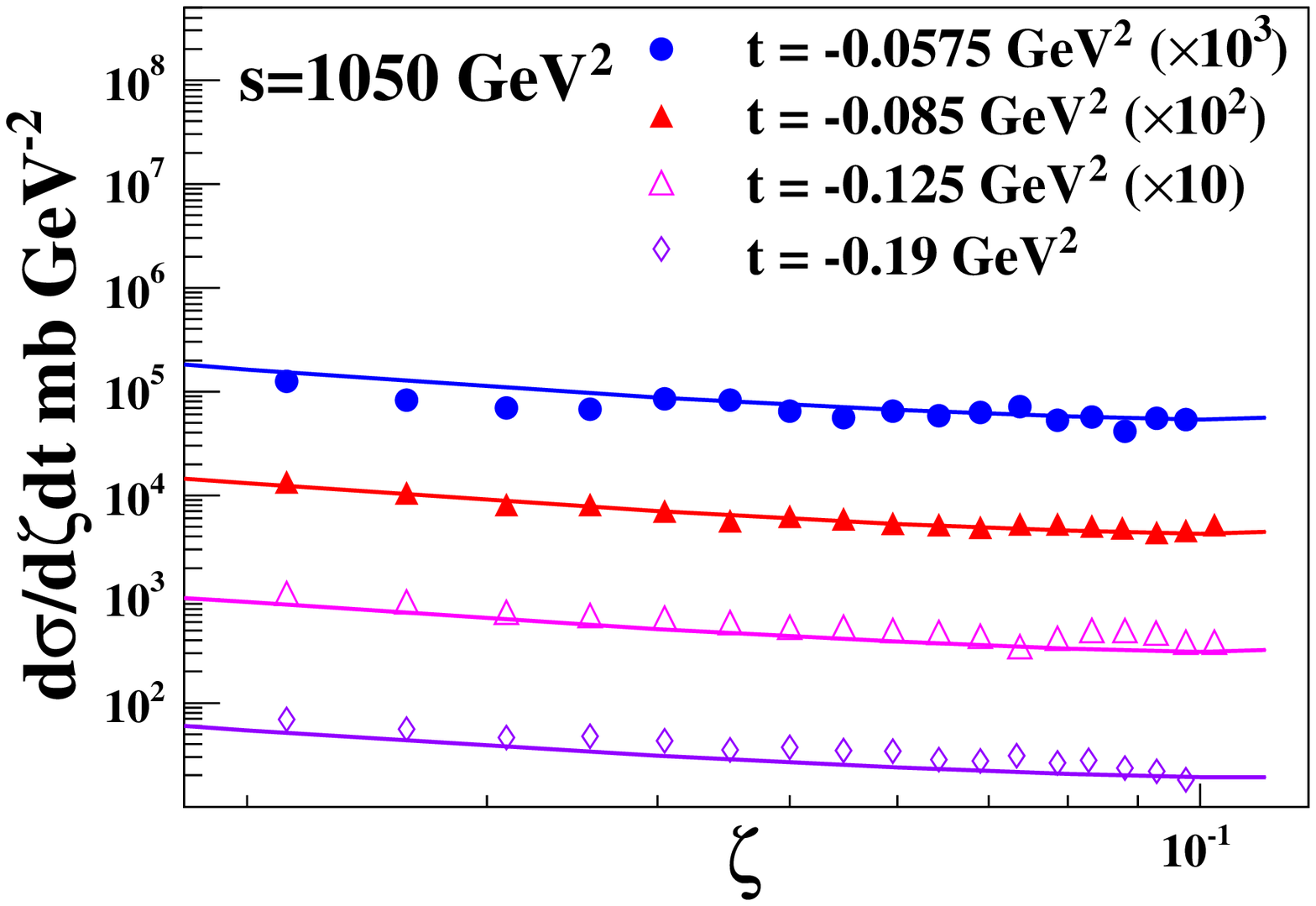}
\end{minipage}
}
\caption{
Double differential cross-section $d^{2}\sigma/d\zeta dt$ for $pp \rightarrow pX$ measured 
at ISR at various $\sqrt{s}$ and t. The data are taken from \cite{Armitage}.}
\label{Fig:Comp3}
\end{figure*}

\vspace{-10pt}
Because we do not consider the contribution of the enhanced diagrams 
 in the elastic scattering amplitude it 
allows us to differentiate data fitting procedure and realise it by 
two steps. At the first step we fix secondary-Reggeon and Pomeron parameters. 
The trajectories of secondary-Reggeons are fixed from fit to data on
spin v.s. mass for corresponding family of mesons and the following results are found:
$\alpha_{f}(t)=0.7+0.8t$, $\alpha_{\omega}(t)=0.4+0.9t$, $\alpha_{\rho}(t)=0.5+0.9t$.  
Then the residues 
of secondary-Reggeons and the residues/trajectory of Pomeron are found 
from fit to data on elastic scattering and total interaction cross-section. 
At the second step we fix triple-Reggeon interaction vertices' constants 
from fit to data on high mass soft-diffraction dissociation, using the values 
of the parameters fixed in the first step as an input.\\ 
We take into account $P$-, $f$- and $\omega$- poles in $pp$ and $p\bar{p}$ elastic scattering amplitude.
Since we assume pion exchange dominance at the coupling of Reggeons, we fix the 
parameters of secondary-Reggeon- and Pomeron- pion coupling 
as well. In $\pi^{\pm}p$ elastic scattering amplitude, 
we take into account $P$-, $f$- and $\rho$- poles. Thus, we assume 
$M=M_P+M_f \pm M_{\omega}$ for $pp$ and $p\bar{p}$ collisions and $M=M_P+M_f \pm M_{\rho}$ 
for $\pi^{+}p$ and $\pi^{-}p$ collisions, respectively. For Pomeron-trajectory we have found
the following parameterization: $\alpha_{P}(t)= 1.117 \pm 0.252t$, and other parameters 
are listed in the Tables~\ref{Tb:ppdata}~and~\ref{Tb:pipdata}.
\begin{table}[h]
\vspace{-10pt}
 \begin{minipage}[c]{0.33\textwidth}
\caption{ $p+p(\bar{p})$ interaction parameters in GeV$^{-2}$ units.\newline
\vspace{-9pt}}
\label{Tb:ppdata} 
\begin{tabular}{l}
\hline
$g_N =  1.366 \pm 0.004 $\\
$R_N^2 =  1.428 \pm 0.006 $\\
$g_N^f =  2.871 \pm 0.008 $\\
$R_N^{f2} = 0.918 \pm 0.023 $\\
$g_N^{\omega} =  2.241 \pm 0.074 $\\
$R_N^{\omega 2} = 0.945 \pm 0.026$\\
\end{tabular}
\end{minipage}
\hspace{10pt}
\begin{minipage}[c]{.3\textwidth}
\caption{$\pi^{\pm}p$ interaction parameters in GeV$^{-2}$ units.\newline
\vspace{-9pt}}
\label{Tb:pipdata}
\begin{tabular}{ll}
\hline
$g_{\pi}$           &\hspace{-10pt}$=0.85 \pm 0.0004$\\
$R_{\pi}^2$         &\hspace{-10pt}$=0.5  \pm 0.002$\\
$g_{\pi N}^f$       &\hspace{-10pt}$=3.524\pm 0.001$\\
$R_{\pi N}^{f2}$    &\hspace{-10pt}$=1.   \pm 0.001$\\
$g_{\pi N}^{\rho}$  &\hspace{-10pt}$=1.12 \pm 0.017$\\
$R_{\pi N}^{\rho 2}$&\hspace{-10pt}$=9.19 \pm 0.837$\\
\end{tabular}
  \end{minipage}
\hspace{10pt}
\begin{minipage}[c]{.25\textwidth}
\caption{Found values of $G_{ijk}$ in GeV$^{-2}$ units.\newline
\vspace{-9pt}}
\label{Tb:Gijk}
\begin{tabular}{ll}
\hline
$G_{PPP}$=0.0098&\hspace{-10pt}$\pm$0.0005\\
$G_{PPR}$=0.03  &\hspace{-10pt}$\pm$0.004 \\
$G_{RRP}$=0.005 &\hspace{-10pt}$\pm$0.001 \\
$G_{RRR}$=0.05  &\hspace{-10pt}$\pm$0.002 \\
$G_{PRP}$=0.013 &\hspace{-10pt}$\pm$0.001 \\
$G_{PRR}$=0.033 &\hspace{-10pt}$\pm$0.005 \\
\end{tabular}
\end{minipage}
\vspace{-8pt}
\end{table}
Next we fix the triple-Reggeon vertices strengths ($G_{ijk}$) from fit 
to data on soft single-diffraction dissociation in $pp$ and $p\bar{p}$ interactions. 
We used the available data on spectra of non-diffracted proton from fixed-target 
experiments \cite{Schamberger} and \cite{Akimov}, from ISR \cite{Armitage} and from CDF \cite{PomFluxRen}. 
Being interested on soft diffraction, we have chosen measurements done for 
$d^2\sigma/d\zeta dt$ within the diffractive cone (-$t \leq$ 0.2 GeV$^2$). 
Found values of $G_{ijk}$ are reported in the Table \ref{Tb:Gijk} and the fit result is compated
with data in Figs~\ref{Fig:Comp1}-\ref{Fig:Comp3}.\\

\begin{figure}[h!]
\vspace{-10pt}
\centering{
\begin{minipage}[l]{0.49\textwidth}
\includegraphics[width=1.\textwidth]{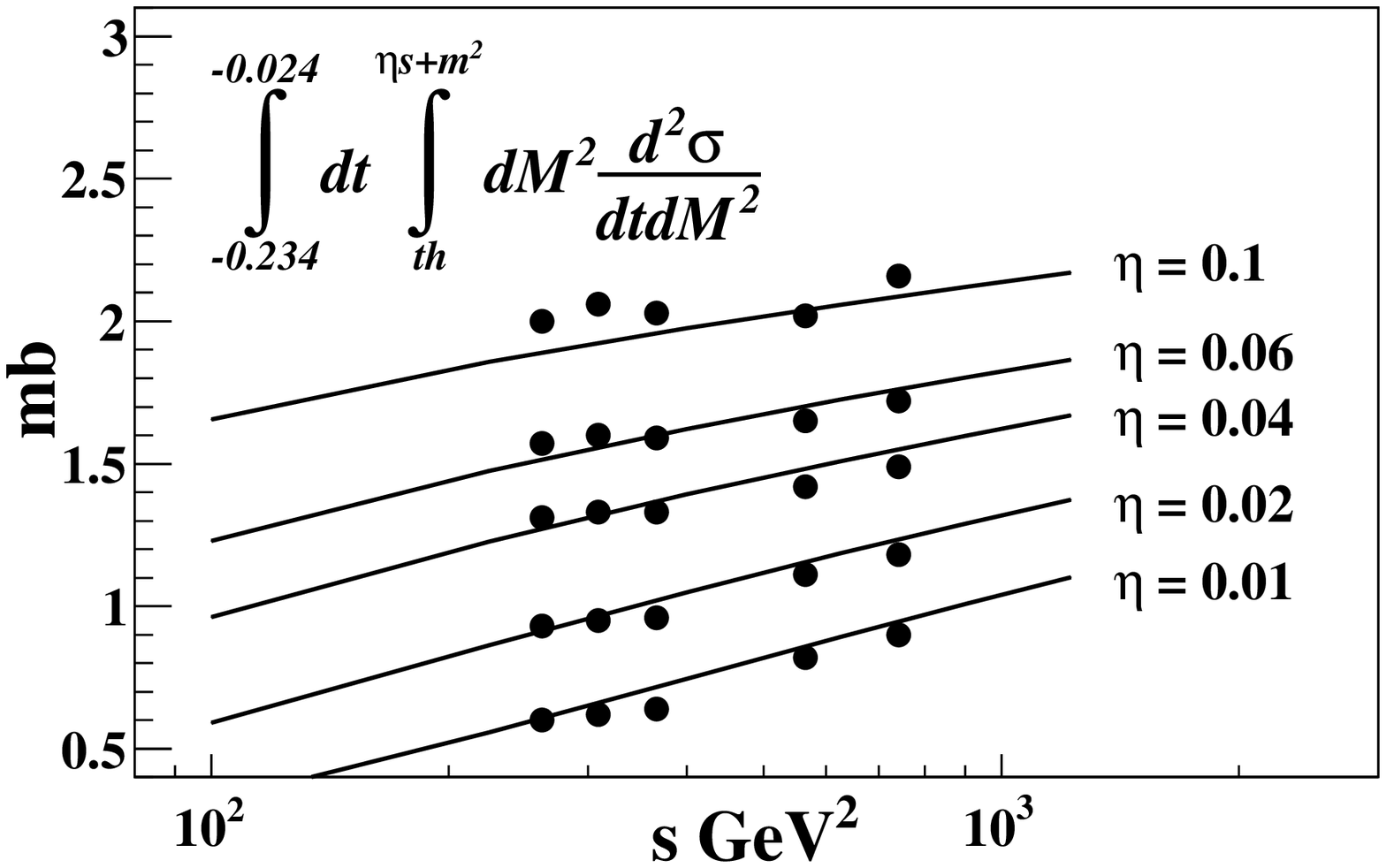}
\end{minipage}
\begin{minipage}[l]{0.49\textwidth}
\includegraphics[width=1.\textwidth]{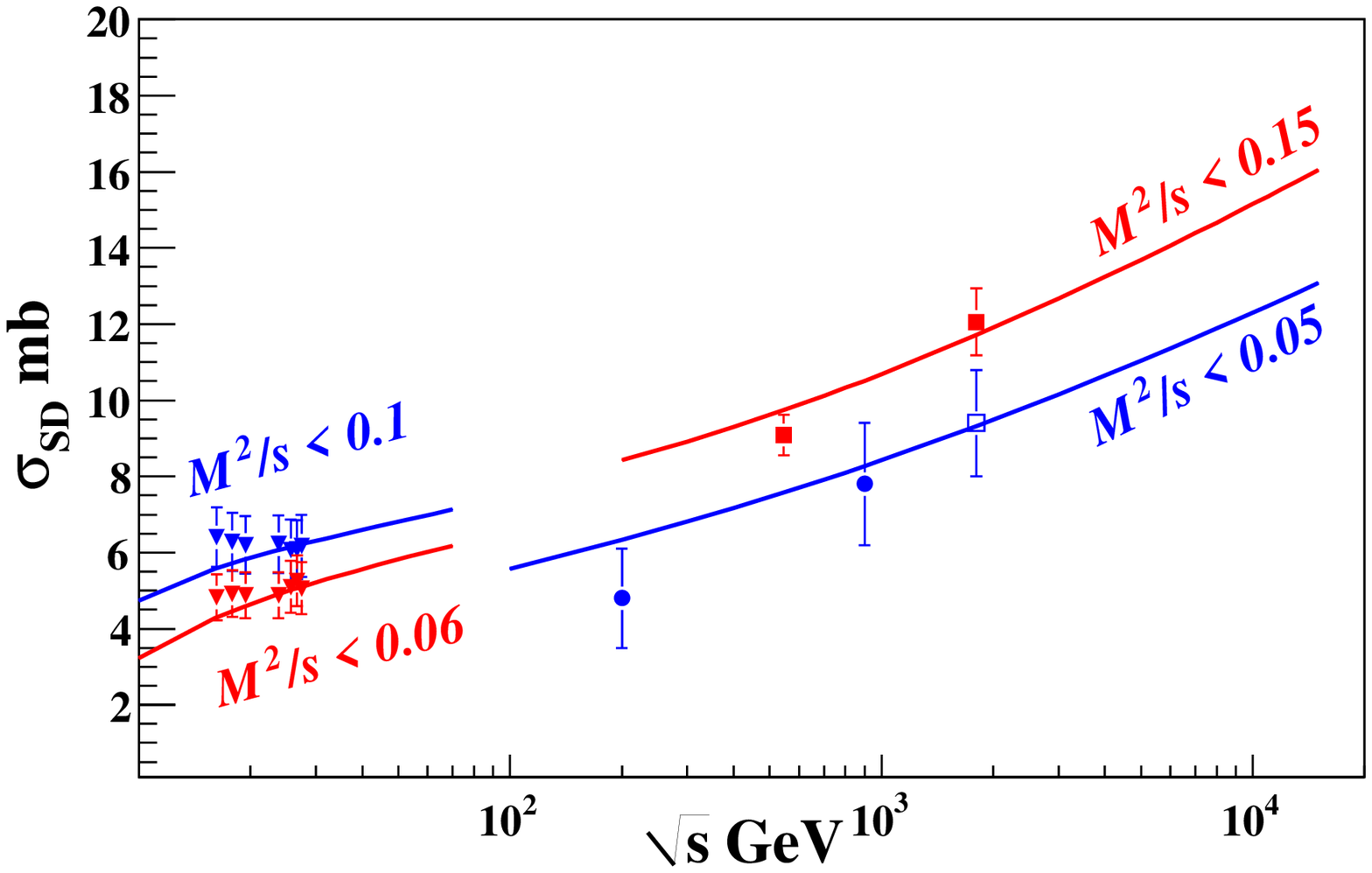}
\end{minipage}
}
\caption{
Integrated single-diffractive cross-section as a function of $\sqrt{s}$. 
The integrations are done in accordance with corresponding measurement as they are 
indicated in the plots. }
\label{Fig:SDcomp}
\vspace{-10pt}
\end{figure}

\begin{SCfigure}[][h!]
  \centering
\includegraphics[width=0.5\textwidth]{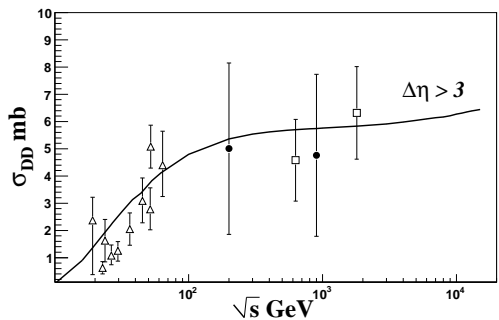}
  \caption{ 
Double-diffractive cross-section as a function of  
$\sqrt{s}$ . The theoretical curve is calculated requiring minimum 3 
units of rapidity gap between two diffracted clusters. }
\label{Fig:DDcomp}
\end{SCfigure}

In Fig.~\ref{Fig:SDcomp} we compare predictions of the model on single-diffractive 
intergrated cross-sections with experimental data \cite{SDXS}. In each case the integration
is done in accordance with the corresponding measurement as they are indicated in the Figure.\\
In Fig.~\ref{Fig:DDcomp} we compare predictions of the model on double-diffractive
intergrated cross-sections with experimental data \cite{Affolder, Alberi}. 
The data at $\sqrt{s}>$100 GeV correspond to the cross-section for minimum 3 units
of rapidity gap between two produced clusters and are taken from \cite{Affolder}.
 The rest of data (at $\sqrt{s}<$100 GeV) are taken from 
\cite{Alberi} where exclusively and semi-inclusively measured data are reduced to totally 
inclusive cross-section. The theoretical curve is calculated using Eq.~(\ref{eq:sigmaDD}) 
and requiring minimum 3 units of rapidity gap between two diffracted clusters.
\vspace{-10pt}
\section{Summary and Predictions for LHC}
\vspace{-5pt}
In this article we report the results of calculations of all non-enhanced absorptive 
corrections to triple-Regge vertices and loop diagrams in eikonal approximation using 
Gribov's Reggeon calculus. Numerically evaluating the model we have found a good 
\begin{wraptable}{r}{0.4\textwidth}
\centerline{
\begin{tabular}{|c|l|l|}
\hline
$\sqrt{s}$ TeV & $\sigma_{SD}$ mb & $\sigma_{DD}$ mb \\   
\hline
0.9            &    8.2           &      5.7         \\
7              &    11.6          &      6.1         \\
10             &     12           &      6.2         \\
14             &     13           &      6.4         \\            
\hline
\end{tabular}}
\caption{Predictions for LHC}
\label{Tb:predictions}
\end{wraptable}
description of data on high-mass soft diffraction dissociation in the energy range from 
ISR, FNAL to Tevatron (from $P_{lab} = 65$ GeV/c  to $\sqrt{s}=1800$ GeV). It is worth 
to emphasize that such a detailed description of inclusive diffraction in this broad 
region of energies is achieved for the first time.\\
In the Table \ref{Tb:predictions} we present the predictions of the model on single- and
double- diffractions cross-section for different energies of LHC.
The single-diffractive cross-section is obtained integrating over masses up to $M^{2}/s=0.05$,
and the double-diffractive cross-section is obtained requiring minimum 3 units of rapidity
gap between two diffracted clusters.\\
We acknowledge A.Grigiryan, J.-P. Revol and K.Safarik for their interest to this work. 
%
%
\begin{footnotesize}

\end{footnotesize}
\end{document}